\documentclass[a4paper,twocolumn,11pt,unpublished]{quantumarticle}
\pdfoutput=1
\usepackage[utf8]{inputenc}
\usepackage[english]{babel}
\usepackage[T1]{fontenc}
\usepackage{amsmath}
\usepackage{hyperref}
\usepackage{dsfont}
\usepackage{subcaption}

\usepackage{tikz}
\usepackage{lipsum}
\usepackage[numbers,sort&compress]{natbib}

\DeclareMathOperator*{\argmin}{argmin}

\newcommand{\ket}[1]{| #1 \rangle}
\newcommand{\bra}[1]{\langle #1 |}
\newcommand{\mean}[1]{\langle #1 \rangle}
\newcommand{\sx}{\hat{\sigma}^x}

\newcommand{\sz}{\hat{\sigma}^z}
\newcommand{\spindown}{\ket{\hspace{-1mm} \downarrow}}
\newcommand{\vtheta}{\boldsymbol{\theta}}
\newcommand{\vbeta}{\boldsymbol{\beta}}
\newcommand{\vgamma}{\boldsymbol{\gamma}}

\newcommand{\UU}{\widehat{U}}
\newcommand{\s}{s}

\newcommand{\cf}{C}

\newcommand{\ffit}{f}
\newcommand{\w}{\boldsymbol{w}}
\newcommand{\ie}{\textit{i.e.} }
\newcommand{\id}{\mathds{1}}
\newcommand{\real}{\operatorname{Re}}
\newcommand{\vecx}{\boldsymbol{\theta}}
\newcommand{\distx}{\xi}
\newcommand{\x}{\theta}

\newcommand{\Xtrain}{\Theta}
\newcommand{\Xtest}{\Theta_*}
\newcommand{\ftrain}{\boldsymbol{f}}
\newcommand{\ftest}{\boldsymbol{f}_*}
\newcommand{\ftestmean}{\overline{f}_*}
\newcommand{\mindist}[1]{\Delta_{min, #1}}
\newcommand{\likelihood}{L}
\newcommand{\noiseless}{}
\newcommand{\GPM}{\ensuremath{{\mbox{\tiny{GPM}}} }}
\newcommand{\Obs}{\hat{O}}
\newcommand{\subspacelabel}{S}
\newcommand{\dimensionlabel}{N}
\newcommand{\nfROTO}{\tilde{n}_f}
\newcommand{\Nsamples}{N_T}

\begin{document}

\title{Gaussian Process Model Kernels for Noisy Optimization in Variational Quantum Algorithms}

\author[1,2]{L. Arceci}
\author[1]{V. Kuzmin}
\author[1,2,3]{R. van Bijnen}
\affil[1]{Institute for Theoretical Physics, University of Innsbruck, Innsbruck A-6020, Austria}
\affil[2]{Institute for Quantum Optics and Quantum Information of the Austrian Academy of Sciences, Innsbruck A-6020, Austria}
\affil[3]{PlanQC GmbH, 85748 Garching, Germany}

\maketitle

\begin{abstract}

Variational Quantum Algorithms (VQAs) aim at solving classical or quantum optimization problems by optimizing parametrized trial states on a quantum device, based on the outcomes of noisy projective measurements. The associated optimization process benefits from an accurate modeling of the cost function landscape using Gaussian Process Models (GPMs), whose performance is critically affected by the choice of their kernel. Here we introduce trigonometric kernels, inspired by the observation that typical VQA cost functions display oscillatory behaviour with only few frequencies. 
Appropriate scores to benchmark the reliability of a GPM are defined, and a systematic comparison between different kernels is carried out on prototypical problems from quantum chemistry and combinatorial optimization. 
We further introduce RotoGP, a sequential line-search optimizer equipped with a GPM, and test how different kernels can help mitigate noise and improve optimization convergence.
Overall, we observe that the trigonometric kernels show the best performance in most of the cases under study. 

\end{abstract}

Variational Quantum Algorithms (VQA) are key candidates for obtaining a quantum advantage with noisy intermediate-scale quantum technologies~\cite{VQA_review_NatRev21, McClean_TheoryVQE_NJP16, VQEreview_chemistry_PhysRep22}. They have been successfully employed for various tasks, including ground state preparation of molecules \cite{Peruzzo_NatComm14, Kandala_Nature17, Roos_Chemistry_PRX18} and quantum many-body systems \cite{Schwinger_Nature19, pagano2020quantum, MethKuzminPRX22}, quantum machine learning~\cite{havlivcek2019supervised, zhu2019training, pan2023experimental}, quantum optimization~\cite{harrigan2021quantum, ebadi2022quantum}, as well as optimal quantum metrology~\cite{Marciniak_sensing_Nature22} and solving linear systems~\cite{VQLS_Quantum23}. 
The core idea involves executing a feedback loop between a quantum device and a classical computer to iteratively solve an optimization problem. 
In each iteration, a parametrized quantum circuit prepares a trial state on the quantum device, which is then probed to estimate an associated cost function value reflecting the quality of the trial state; based on this value (and possibly previous outcomes), a classical computer then optimizes the circuit parameters for the next iteration.

Finding the global minimum of a high-dimensional optimization problem is generally challenging. In VQA problems, this task becomes even more difficult due to quantum projection noise (measurement noise) obscuring the true cost function values. 
One way to ameliorate this problem is to construct a surrogate model \cite{McClean_surrogateModels_QST20, FontanaRunngerCirstoiu_arXiv22, Eisert_surrogateModels_PRL23, SGLBO_NPJ22, koczor2022quantum, stanisic2022observing},
incorporating all available measurement data. 
This model can then be used either to provide better estimates of the true cost function value or to make inferences on unmeasured points. 
In this setting, Gaussian Process Models (GPM) \cite{Rasmussen_GPM_book} are an effective strategy for building reliable surrogate models and mitigating the effect of noise during the optimization. 
They have been introduced as a tool to assist optimization in VQA in different ways, both for global \cite{Jones_BayesianOpt, Schwinger_Nature19, Marciniak_sensing_Nature22, Ercolessi_BayesianOpt_QAOA} and local optimizers \cite{SGLBO_NPJ22, stanisic2022observing}. 

\begin{figure}
  \centering
  \includegraphics[width=\columnwidth]{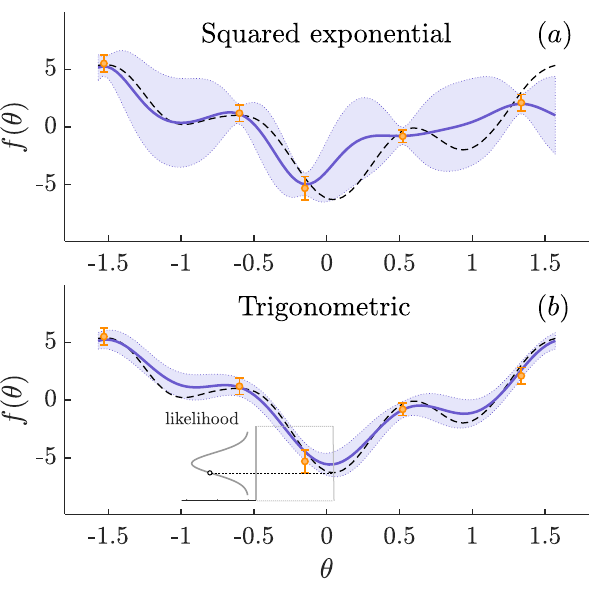}
  \caption{GPM fits of a noisy data set (orange dots) using (a) squared exponential (Eq.~\eqref{eq:kernelSE}) and (b) trigonometric kernel (Eq.~\eqref{eq:kernel:TRIG}). The blue solid line is the average GPM prediction and the shaded area corresponds to one $\sigma$ of uncertainty; the black dashed line is the exact cost function.
  Inset in (b): GPM-predicted probability density at the true minimum, yielding the likelihood. 
  }
  \label{fig:visualizeGPM}
\end{figure}

In this work, we investigate the role of the \textit{kernel} function \cite{Rasmussen_GPM_book} of a GPM, which characterizes the correlations between cost function values at different circuit parameters. The choice of the kernel function is crucial as it directly influences the model's ability to capture complex patterns in the data, its generalization performance, and its predictive uncertainty estimation (see Fig.~\ref{fig:visualizeGPM}). Despite its importance, very little attention has been given to which kernel choices are best suited for tackling VQA problems. Here we address this issue by systematically investigating the performance of various kernels, including kernels tailored specifically to VQAs, that we call \textit{trigonometric kernels}.

We compare the goodness of GPM fits using different data collections, coming from two prototypical VQA problems: finding the ground state of the LiH molecule with Unitary Coupled-Cluster Ansatz truncated to Single and Double excitations (UCCSD)~\cite{Kandala_Nature17, Roos_Chemistry_PRX18, VQEreview_chemistry_PhysRep22} and solving instances of unweighted MaxCut problem~\cite{Lucas_IsingOptimizations} with the Quantum Approximate Optimization Algorithm (QAOA)~\cite{Fahri_QAOA_arXiv14}. 
These problems exhibit a phenomenon noted before in VQAs, namely that the cost function shows oscillatory behavior with only few dominant frequencies (see Ref.~\cite{Eisert_surrogateModels_PRL23, FontanaRunngerCirstoiu_arXiv22, FastGradient_Bittel_arXiv22} and Appendix~\ref{app:fewfreqs}), far fewer than one would a priori expect~\cite{NFT_PRR20}. 
This observation is crucial to construct the trigonometric kernels. 
As a final application, we furthermore introduce a GPM-enhanced sequential line optimizer~\cite{SLS_Parrish_arXiv19, NFT_PRR20, Rotosolve_Quantum21}, RotoGP, whose name takes inspiration from the RotoSolve optimizer~\cite{Rotosolve_Quantum21}. We evaluate its performance using different kernels and show that the trigonometric one provides the best results overall.

\section{Kernels for Gaussian Process Models}
\label{sec:KernelsIntro}
In the following we introduce Gaussian Process Models (GPM) of VQA cost functions. After a brief introduction to the subject in Sec.~\ref{sec:KernelsIntro:GPM}, we review some of the most common kernels used in the literature in Sec.~\ref{sec:KernelsIntro:reviewKernels} and subsequently derive the trigonometric kernel specialized to VQAs in Sec.~\ref{sec:KernelsIntro:TRIGkernel}. 
Finally, in Sec.~\ref{sec:KernelsIntro:prodKernels}, we explain how to construct valid kernels in high-dimensional spaces by using products of one-dimensional kernels. 

\subsection{Gaussian Process Models of VQA cost functions}
\label{sec:KernelsIntro:GPM}
In Variational Quantum Algorithms (VQA) \cite{McClean_TheoryVQE_NJP16, VQA_review_NatRev21}, 
starting from an initial state $\ket{\psi_0}$, a parameterized quantum circuit $\UU(\vecx)$ prepares a state
$\ket{\psi(\vecx)} = \UU(\vecx) \ket{\psi_0}$ 
on the quantum device, where $\vecx$ is a vector of control parameters. 
The system is then measured to estimate expectation values of $N_O$ observables $\{\widehat{O}_i\}_{i = 1}^{N_O}$ on $\ket{\psi(\vecx)}$, which are used to compute a cost function value $\cf(\vecx)$ to be minimized,
\begin{equation}
\cf(\vecx) = \bra{\psi(\vecx)} \sum_i \widehat{O}_i \ket{\psi(\vecx)}.
\label{eq:VQA:costfunction}
\end{equation}
Accurate estimation of the expectation value of each $\widehat{O}_i$ requires several runs of the same circuit 
--- called \textit{shots} --- 
to collect a set of projective measurement values $\{o_{i,j}\}_{j=1}^s$ to average over,
\begin{equation}
    \bra{\psi(\vecx)} \widehat{O}_i \ket{\psi(\vecx)} \approx \frac{1}{\s} \sum_{j=1}^\s o_{i,j} \; ,
    \label{eq:VQA:finiteSampling}
\end{equation}
where $\s$ is the number of shots used. 
Given that $\ket{\psi(\vtheta)}$ is generally not a simultaneous eigenfunction of all $\widehat{O}_i$, the values $o_{i,j}$ will fluctuate, endowing the cost function with inherent quantum projection noise and its estimated value with an inherent error bar (see Fig.~\ref{fig:visualizeGPM}). Under relatively general assumptions, this noise is Gaussian distributed around the true cost function value~\cite{McClean_TheoryVQE_NJP16} (although we will discuss an exception for the MaxCut problem in Sec.~\ref{sec:KernelComparison:Minimum}).

Gaussian Process Modelling is a machine learning technique used for regression or classification of (noisy) data sets \cite{Rasmussen_GPM_book}. Here, one models a (cost) function as a Gaussian process, which is formally defined as a collection of random variables, any finite number of which have a joint Gaussian distribution~\cite{Rasmussen_GPM_book}. That is, the cost function values $\cf(\vecx_1), \cf(\vecx_2), \dots, \cf(\vecx_{\Nsamples})$ at a collection of points $\{\vecx_i\}_{i=1}^{\Nsamples}$ in the parameter space are modeled as Gaussian random variables $f_1, f_2, \dots f_{\Nsamples}$, with mean values $\mu_i$ which represent our prior knowledge of the function value, \ie what function value we would expect at a point $\vecx_i$ before sampling anywhere. 
The joint distribution of these Gaussian random variables is furthermore characterized by covariances, expressed through a \textit{kernel function} $k(\vecx_i, \vecx_j)$, quantifying the expected correlations between cost function values at different points $\vecx_i$ and $\vecx_j$. 
If the covariance $k(\vecx_i, \vecx_j)$ is large and positive (negative), fluctuations of the random variables around their mean at points $\vecx_i$ and $\vecx_j$ are correlated (anticorrelated). If the covariance is zero, function values at points $\vecx_i$ and $\vecx_j$ are uncorrelated, \ie knowledge about the function value at one point does not affect our estimates at the other point in any way. 
The kernel function $k(\vecx_i, \vecx_j)$ encodes precisely how function values at distinct points are correlated through its functional dependence on $\vecx_i$ and $\vecx_j$, and we will review specific example kernels in the next subsection.

When employing a GPM in practice, for a given choice of kernel, one starts with a training set of $\Nsamples$ points $\Xtrain = \{\vecx_i\}_{i = 1}^{\Nsamples}$ in parameter space 
with associated observed values $\mathbf{y} = \{y_i\}_{i = 1}^{\Nsamples}$
and (potentially) error bars $\boldsymbol{\sigma} = \{\sigma_i\}_{i = 1}^{\Nsamples}$; 
then, the goal is to make a prediction on the associated values at another set of points in parameter space, which constitutes a test set $\Xtest$. 
Given $\ftrain$ ($\ftest$) as the vector of GPM random variables for training (test) points, the prior multivariate Gaussian distribution is then given as 
\begin{equation}
    \begin{bmatrix}
        \ftrain \\
        \ftest
    \end{bmatrix}
    \sim \mathcal{N} \left(
    \boldsymbol{0}, 
    \begin{bmatrix}
        \Tilde{K}_{\Xtrain,\Xtrain} & K_{\Xtrain,\Xtest} \\
        K_{\Xtest,\Xtrain} & K_{\Xtest,\Xtest}
    \end{bmatrix}    
    \right) \; ,
    \label{eq:GPMprior}
\end{equation}
where the covariance matrices $K_{\Xtrain, \Xtrain'}$ are evaluated from the kernel function
\begin{equation}
\begin{bmatrix}
K_{\Xtrain, \Xtrain'}
\end{bmatrix}_{i, j} = k(\vecx_i, \vecx_j')
\end{equation}
and where 
\begin{equation}
\Tilde{K}_{\Xtrain,\Xtrain} = K_{\Xtrain,\Xtrain} + \mathrm{diag}(\{\sigma^2_n\}_{n = 1}^{N}),
\end{equation}
\ie we add the observed variances in the diagonal of the training covariance matrix $K_{\Xtrain,\Xtrain}$; in doing this, we assume that the measurement noise is additive and follows an independent, Gaussian distribution with zero mean. 
Note that in Eq.~(\ref{eq:GPMprior}), we have also set the prior mean values to zero, $\boldsymbol{\mu} = \boldsymbol{0}$, which can be done without loss of generality and is common in GPM literature~\cite{Rasmussen_GPM_book}. Moreover, in the case of VQA, the typical cost function is a weighted sum of Pauli strings, each of which has expectation value zero for an a-priori unknown state.

We subsequently condition the distribution in Eq.~\eqref{eq:GPMprior} on the observations $(\mathbf{y}, \boldsymbol{\sigma})$ at the training points $\Xtrain$. The posterior distribution is again multivariate Gaussian~\cite{Rasmussen_GPM_book}, but with new mean and covariance that depend on the observations and on the kernel function,
\begin{align}
\begin{split}
    \ftestmean &= K_{\Xtest,\Xtrain} \Tilde{K}_{\Xtrain,\Xtrain}^{-1} \mathbf{y} \\
    \mathrm{cov}(\ftest) &= K_{\Xtest,\Xtest} - K_{\Xtest,\Xtrain} \Tilde{K}_{\Xtrain,\Xtrain}^{-1} K_{\Xtrain,\Xtest} \; .
\end{split}
\label{eq:GPMposterior}
\end{align}
Thus, $\ftestmean$ is the vector of GPM predictions, with corresponding variances given in the diagonal of $\mathrm{cov}(\ftest)$. 

An important feature of GPMs is that one can employ hyperparameters $\{\boldsymbol{\alpha}\}$ in the kernel function, such that the model becomes more flexible. The GPM framework allows to automatically optimize these hyperparameters using the data at hand~\cite{Rasmussen_GPM_book}. To do so, 
one can compute the likelihood of the observations $\mathbf{y}$, under the GPM probability distribution, conditioned on the training points $\Xtrain$ and the model's hyperparameters, 
\begin{equation}
    p(\mathbf{y}|\boldsymbol{\alpha}, \Xtrain) = 
    \frac{\exp \left( -\frac{1}{2} \mathbf{y}^T ( \Tilde{K}_{\Xtrain,\Xtrain}(\boldsymbol{\alpha}) )^{-1} \mathbf{y} \right)}
    {\sqrt{(2\pi)^{\Nsamples} \det\left( \Tilde{K}_{\Xtrain,\Xtrain}(\boldsymbol{\alpha}) \right) }} \; ,
    \label{eq:GPMlikelihoodHPs}
\end{equation}
and find the hyperparameters that maximize it \cite{Rasmussen_GPM_book}. Importantly, the likelihood formula reflects an inherent trade-off between the model's accuracy and complexity. Hyperparameters that yield high agreement with the training data result in high values of the numerator, promoting the accuracy of the model. However, this agreement may come at the cost of increased model complexity, which is penalized by the determinant term in the denominator. This balance prevents overfitting, encouraging simpler models that generalize better.

We note that the whole GPM framework, as discussed above, can be more formally derived from a fitting of basis functions (regression) under certain assumptions of Gaussianity and Bayesian updating of fitting weights~\cite{Rasmussen_GPM_book}. 
More precisely, one can identify a set of basis functions $\boldsymbol{\phi}(\vecx) = \{\phi_1(\vecx), \phi_2(\vecx), \ldots \}$ such that predictions of the cost function values follow the equation 
\begin{equation}
    \ffit(\vecx) = \w \cdot \boldsymbol{\phi}(\vecx) \; ,
    \label{eq:wdotphi}
\end{equation}
where $\w$ is a vector of stochastic variables, obeying a zero-mean Gaussian distribution with prior covariance $\Sigma_p$, \ie $\w \sim \mathcal{N}(\boldsymbol{0}, \Sigma_p)$. 
It can be shown that the set of variables $\ftrain(\vecx)$ from this approach forms a Gaussian process~\cite{Rasmussen_GPM_book}. Therefore, all the previous discussion applies. 
Importantly, taking this approach, one can derive the kernel associated with the GPM directly from the basis functions $\boldsymbol{\phi}(\vecx)$ and the prior on the weights covariance $\Sigma_p$. 
This is the path we will follow to derive the trigonometric kernel in Sec.~\ref{sec:KernelsIntro:TRIGkernel}.

\subsection{Review of standard kernels}
\label{sec:KernelsIntro:reviewKernels}
The most popular and widely used kernels are so-called stationary kernels. 
This class of functions depends only on the distance between points and not on their individual values, 
\ie $k(\vecx, \vecx') = k(\distx)$ with $\distx = |\vecx-\vecx'|$. 
In the following, we introduce each kernel in a one-dimensional setting, \ie the parameter vector $\vecx = \x$ is a scalar, and $\xi = |\x-\x'|$. A convenient generalization to higher dimensions is discussed in Sec.~\ref{sec:KernelsIntro:prodKernels}.

Let us start with the \textit{Squared-Exponential} (SE) kernel \cite{Rasmussen_GPM_book}, 
\begin{equation}
    k_{SE}(\distx) = \exp \left(-\frac{\distx^2}{2l^2} \right) \; ,
    \label{eq:kernelSE}
\end{equation}
where $l$ is a hyperparameter controlling the correlation length of points in the cost function that essentially defines how stiff or flexible the GPM fit is. 
Functions drawn from a Gaussian process with a SE kernel are infinitely differentiable and, therefore, smooth. 

Sometimes, however, one might need to relax the smoothness assumption slightly. In these cases, the \textit{Mat\'ern} family of kernels \cite{Rasmussen_GPM_book} can be a better choice. The most used instances are 
\begin{align}
    &k_{M,3/2}(\distx) = \left( 1 + \frac{\sqrt{3}\distx}{l} \right) \exp \left(-\frac{\sqrt{3}\distx}{l} \right) \\
    &k_{M,5/2}(\distx) = \left( 1 + \frac{\sqrt{5}\distx}{l} + \frac{5\distx^2}{3l^2} \right) \exp \left(-\frac{\sqrt{5}\distx}{l} \right) \; ,
\end{align}
again with $l$ as hyperparameter. A Gaussian process with these kernels produces functions that are only one and two times differentiable, respectively. They can, therefore, exhibit a less smooth behavior than the squared exponential kernel. 

In many cases of interest for VQA, we know that the cost function is periodic, at least along a subset of parameters. In these cases it would be desirable to encode this information in the kernel. 
To this purpose, the \textit{periodic SE} kernel (pSE) is an interesting option \cite{Rasmussen_GPM_book}, 
\begin{equation}
    k_{pSE}(\distx) = \exp \left( -\frac{1}{2 l^2} 
                \left[ \frac{p}{\pi}  
                \sin \left( \frac{\pi \distx}{p} \right) \right]^2
                \right) \; ,
    \label{eq:kernelSEperiodic}
\end{equation}
where $p$ is the period of the cost function, and $l$ is again a hyperparameter. 
Note that, in the limit of $\distx \ll p/\pi$, this kernel reduces locally to a standard SE. 

\subsection{The trigonometric kernel}
\label{sec:KernelsIntro:TRIGkernel}
In many VQA applications, one can put more insight into the GPM kernel. 
Along with periodic dimensions, cost functions can, in general, be written as a linear combination of trigonometric functions with frequencies that are an integer multiple of a base frequency. We illustrate this with a simple example for two qubits: let us assume that 
the variational state is prepared using a single parametrized gate as 
$\ket{\psi(\x)} = \exp(-\mathrm{i} \x [\sx_1 + \sx_2]) \ket{\psi_0}$. 
Such a gate would, for instance, occur as the driver Hamiltonian in QAOA~\cite{Fahri_QAOA_arXiv14}. Using that $\exp(-\mathrm{i} \x \sx) = \cos\x - \mathrm{i} \sin\x \: \sx$, and writing $\Obs = \sum_i\hat{O}_i$, we have for our cost function
\begin{equation}
    \bra{\psi(\x)} \Obs\ket{\psi(\x)} = \sum_{ \substack{p, q, r, s \\ \in \{0, 1\}} }  c_{pqrs} \; \cos^\alpha \x \; \sin^{4-\alpha} \x \; ,
    \label{eq:costfunexpansion}
\end{equation}
with 
$c_{pqrs} = (-i)^\alpha (-1)^{r+s} \mean{ \psi_0 | \hat{\sigma}_1^p \hat{\sigma}_2^q \Obs \hat{\sigma}_1^r \hat{\sigma}_2^s | \psi_0 }$, 
$\hat{\sigma}_i^0 = \id_i$, $\hat{\sigma}_i^1 = \hat{\sigma}_i^x$ and
$\alpha = p+q+r+s$.
Clearly, the cost function is simply a \textit{sum of trigonometric functions} times some static expectation values.
The above example can be easily generalized to more qubits and deeper circuits. The cost function will remain a sum of trigonometric functions for each parameter for which the applied gate operators have regularly spaced eigenvalues, as is the case for all (products of) Pauli operators with eigenvalues $\pm1$. Even operators that do not have regularly spaced eigenvalues might still be well-approximated with a trigonometric series expansion. 

Note that one would, in principle, expect that the number of trigonometric functions appearing in the cost function grows exponentially in the number of parameters and qubits. 
However, in practice, it was observed that cost functions from QAOA for combinatorial optimization problems exhibit much fewer relevant frequencies in their spectra than expected~\cite{Eisert_surrogateModels_PRL23, FontanaRunngerCirstoiu_arXiv22, FastGradient_Bittel_arXiv22}. 
In Appendix \ref{app:fewfreqs}, we verify that this phenomenon is present for the test problems that we consider here by analyzing several instances of QAOA on the MaxCut problem, as well as a prototypical instance of quantum chemistry problem with UCCSD ansatz. 
We find that along coordinate lines, \ie, scanning the cost function letting only one parameter vary, only a few of the lowest frequency components play a role, while the others either are zero or decay exponentially in magnitude. 

Based on this insight, we propose a new kernel that we call the \textit{trigonometric kernel}. 
Rather than relying on very general properties (such as smoothness and differentiability), we start its derivation by identifying a set of trigonometric basis functions $\boldsymbol{\phi}(\x)$ that describe the (periodic) cost function along a coordinate line. Then, we will use the same basis functions to define a GPM and construct the associated kernel. 
In the following, we assume that the cost function can be written as a trigonometric polynomial 
\begin{align}
    \cf(\x) = a_0 + \sum_{n=1}^M \Big( a_n \cos(n\x) + b_n \sin(n\x) \Big) \; ,
\end{align}
with coefficients 
$\boldsymbol{a} = [a_0, a_1, ..., a_M]$, 
$\boldsymbol{b} = [b_1, ..., b_M]$, 
$\x \in [0, 2\pi]$, and with a maximum number of frequencies $M$. We identify the vector of basis functions as 
\begin{align}
    \boldsymbol{\phi}(\x) = 
    \begin{bmatrix}
        1 \\
        \cos \x \\
        ... \\
        \cos (M\x) \\
        \sin \x \\
        ... \\
        \sin (M\x)
    \end{bmatrix} \; ,\label{eq:basisfunctionsvector}
\end{align}
such that 
$\cf(\x) = [\boldsymbol{a}, \boldsymbol{b}] \cdot \boldsymbol{\phi}(\x)$. 
The expansion in basis functions is, in principle, exact. However, the coefficient vector is unknown. Finding (an efficient description of) the coefficients would constitute a successful modeling of the cost function, and in the following, we will use a GPM to do so. 

We start by treating the vector of unknown coefficients $\w \equiv [\boldsymbol{a}, \boldsymbol{b}]$ as Gaussian random variables, with zero mean $\mean{\w} = 0$, and covariance matrix $\Sigma_p$, 
as detailed in Sec.~\ref{sec:KernelsIntro:GPM}, and consider the GPM resulting from Eq.~\eqref{eq:wdotphi}. 
The associated covariance kernel follows directly from the basis functions and the covariance matrix~\cite{Rasmussen_GPM_book}, 
\begin{equation}
    k(\x, \x') = \boldsymbol{\phi}^T(\x) \cdot \Sigma_p \cdot \boldsymbol{\phi}(\x') \; .
    \label{eq:kernel_basis}
\end{equation}
In principle, this directly suggests a new kernel, with $M^2$ hyperparameters gathered in the matrix $\Sigma_p$. However, in practice, there are too many hyperparameters, and some simplifying assumptions are required. 

First, we assume no correlations between weights, i.e., we set the off-diagonal elements of $\Sigma_p$ to zero. Next, we leverage the aforementioned observation that only the lowest few frequencies appear to be relevant in the cost function landscape in practice. We incorporate this knowledge in the kernel by retaining only those hyperparameters corresponding to the lowest few frequencies. The remainder of the coefficients $\w$ can be assumed either to be exactly zero or to decay exponentially. 
We follow here the first approach, leaving the derivation using the second option to Appendix \ref{app:trigDetails}. 
We therefore set $[\Sigma_p]_{ii} = 0$ for all the frequencies $n_i > n_f$, with $n_f$ the threshold frequency, and we arrive at the following, \textit{non-stationary trigonometric kernel} (nsTRIG):
\begin{align}
\begin{split}
    k_T^{(ns)}(\x, \x') = \Bigg\{ \gamma_0 + \sum_{n=1}^{n_f} &\Big[ \gamma_n \cos (n\x) \cos (n\x') \;  \\
                + &\delta_n \sin (n\x) \sin (n\x') \Big] \Bigg\} / \Gamma,
\end{split}
\label{eq:kernel:TRIG}
\end{align}
where $\gamma_n$, $\delta_n$ are hyperparameters, and the factor $\Gamma = \gamma_0 + \frac{1}{2} \sum_{n=1}^{n_f} \left( \gamma_n + \delta_n \right)$ serves to normalize the kernel on average over the whole period, that is 
$\langle k(\x,\x) \rangle_{[0, 2\pi]} = 1$. 
An additional overall amplitude hyperparameter will then be included once we generalize the kernel to higher dimensions in the next section.
Note that the kernel~\eqref{eq:kernel:TRIG} is described overall by $2n_f + 1$ hyperparameters. 

We can also obtain a stationary version of this kernel. If we further assume that $\gamma_n = \delta_n$ for all $n \geq 1$, we arrive at the \textit{stationary trigonometric kernel} (sTRIG),
\begin{equation}
    k_{T}^{(s)}(\distx) = \Bigg\{ \gamma_0 + 
    \sum_{n=1}^{n_f} \gamma_n \cos \left( n \distx \right) 
    \Bigg\} / \Gamma,
\label{eq:kernel:TRIGstat}
\end{equation}
where $\distx = |\x-\x'|$ and we employ $n_f + 1$ hyperparameters.

\subsection{Product kernels for high-dimensional spaces}
\label{sec:KernelsIntro:prodKernels}
The kernels discussed thus far were introduced in a one-dimensional setting, \ie employing
\textit{line kernels} for modeling the cost function along a coordinate axis (line) in parameter space. 
We discuss here a way to generalize them to higher-dimensional spaces by using a product of different line kernels, one for each parameter, since a product of valid kernels is also a valid kernel \cite{Rasmussen_GPM_book}. 
Therefore, we construct a kernel for a $\dimensionlabel$-dimensional space as 
\begin{equation}
    k_j^{(\dimensionlabel)}(\vecx, \vecx') = A \prod_{n=1}^\dimensionlabel k_{j,n}(\x_n, \x'_n) \, ,
    \label{eq:kernelProduct}
\end{equation}
where $j$ stands for any of the previous line kernel labels, $A$ is an additional hyperparameter that controls the variance of the prior probability distribution at each point, and each line kernel $k_{j,n}(\x_n, \x'_n)$ has its own hyperparameters and corresponds to the $n^{th}$ dimension ($n^{th}$ parameter). 
Note that kernel in Eq.~\eqref{eq:kernelProduct} can be written in the form of Eq.~\eqref{eq:kernel_basis} for the multi-parameter case, using a prior in the form 
$\Sigma_p = \bigotimes_i \Sigma_p^{(i)}$, where $\Sigma_p^{(i)}$ is the prior for the $i^{th}$ line kernel, 
and using the basis functions $\boldsymbol{\phi}(\vecx) = \bigotimes_i \boldsymbol{\phi}(\x_i)$.

\begin{figure*}
  \centering
  \includegraphics[width=\textwidth]{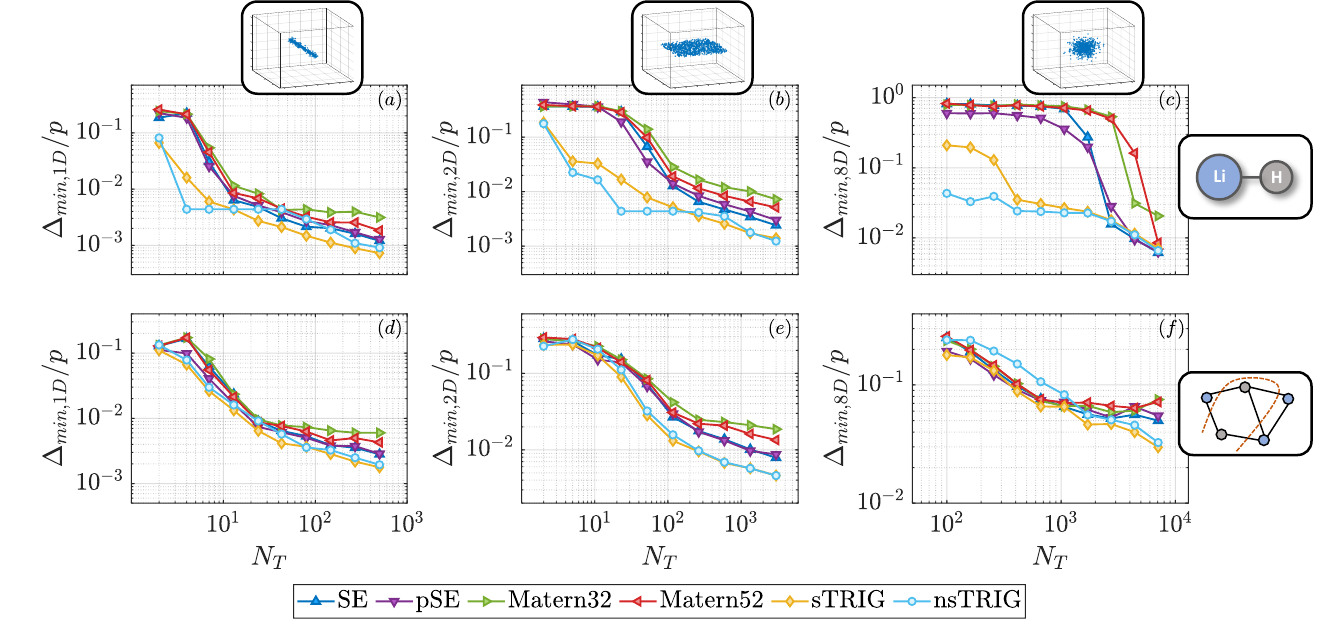}
  \caption{Distance between GPM-predicted and exact global minimum points (rescaled to the cost function period $p$), for different kernels and sizes of the training set $\Nsamples$, for the quantum chemistry (top row) and the combinatorial optimization problems (bottom row). 
  Panels on each column show results for training data lying on the different subspaces of the parameter space described in Sec.~\ref{sec:KernelComparison:HamiltoniansAndSampling}.  
  Each point is the median over 100 values obtained from different training sets of the same size, common to all the kernels. 
  }
  \label{fig:KernelComparison:MinimumPoint}
\end{figure*}

\section{Comparison of Kernels}
\label{sec:KernelComparison}
In this Section, we compare the GPM fitting capabilities for the different kernels introduced in Sec.~\ref{sec:KernelsIntro} and under different scenarios. 
In Sec.~\ref{sec:KernelComparison:HamiltoniansAndSampling}, we describe the two prototypical VQA problems considered for the benchmark, and we explain how we build different training sets of points in parameter space. 
In the subsequent sections, we assess the reliability of each kernel in 
1) locating the global minimum point (Sec.~\ref{sec:KernelComparison:MinimumPoint}), 
2) predicting the correct minimum cost function value (Sec.~\ref{sec:KernelComparison:Minimum}) and 
3) fitting generic points in the training subspace (Sec.~\ref{sec:KernelComparison:TestSet}).

\subsection{Problem Hamiltonians and sampling strategies}
\label{sec:KernelComparison:HamiltoniansAndSampling}
In this section, we introduce the different settings under which the benchmark will be carried out. 
We focus on two types of cost functions, coming from different prototypical VQA problems. 
The first concerns the ground state search of the LiH molecule using one layer of UCCSD Ansatz \cite{Roos_Chemistry_PRX18, VQA_review_NatRev21, VQEreview_chemistry_PhysRep22}; 
the second is the unweighted MaxCut combinatorial optimization problem \cite{Lucas_IsingOptimizations} on 3-regular graphs with $8$ vertices, whose solution is sought using the Quantum Approximate Optimization Algorithm (QAOA) with $4$ layers \cite{Fahri_QAOA_arXiv14}. 
These problems are described in detail in Appendix~\ref{app:LiH_intro} and \ref{app:uwMaxCut_intro}, respectively. Both problems require $N=8$ circuit parameters, and we use $\s = 20$ shots to estimate the expectation value of each observable $\mean{ \widehat{O}_i }$, see Eq.~\eqref{eq:VQA:finiteSampling}. 
We performed the simulations using the Qiskit and Qiskit-Nature packages~\cite{qiskit, qiskit_nature}. 

The corresponding cost functions have a feature in common: they are periodic and described by only the lowest frequencies, as shown in Appendix~\ref{app:fewfreqs}.
The LiH problem is very well described by only the first two frequencies and a period of $p = 2\pi$, along every coordinate line; the unweighted MaxCut problem is instead well approximated by the lowest four (the lowest) frequencies along odd (even) coordinates, with a period $p = \pi$. 
We therefore set frequency thresholds $n_f = 2$ for trigonometric kernels along all directions for the chemistry problem, and $n_f = 4$ ($n_f = 1$) for odd (even) coordinate lines for combinatorial optimization.

We consider three different distributions of the sampling data fed into the GPM: 
\begin{enumerate}
\item  samples randomly distributed around the line parallel to the $5^{th}$ coordinate (corresponding to a parameter in the middle of the circuit) and passing through the global minimum 
$\vecx_{min}^{\noiseless}$, 
with distances from the line uniformly distributed in the range $[0, 0.05\pi]$; 
\item  samples randomly distributed around the plane spanned by the $4^{th}$ and $5^{th}$ coordinates and passing through $\vecx_{min}^{\noiseless}$, 
with distances from the plane uniformly distributed in the range $[0, 0.05\pi]$; 
\item  samples uniformly distributed in a 8D hyperball centered at $\vecx_{min}^{\noiseless}$ with radius $r = 0.15\pi$. 
\end{enumerate}
The latter scenario mimics the later stages of a local optimization, where the optimization algorithm is probing points in the vicinity of the true minimum. 
The other two scenarios focus instead on data approximately lying on subspaces of small dimension, where GPM are expected to work better in general~\cite{Kandasamy_BO_highD, SGLBO_NPJ22}.
Each sampling strategy provides data from within a specific subspace $\subspacelabel$ of the total parameter space. To refer to them, we will use the labels $\subspacelabel = $ 1D, 2D and 8D for sampling strategies 1, 2 and 3, respectively. 

For each sampling strategy and for a given number of training samples $\Nsamples$, we build one hundred training sets with points selected at random (according to the selected sampling strategy). 
In the unweighted MaxCut case, each training set at fixed $\Nsamples$ is taken from a different graph, \ie a different problem instance. 
Given a training set, we first train a GPM by optimizing its hyperparameters following the standard strategy of maximizing the likelihood on the training data from Eq.~\eqref{eq:GPMlikelihoodHPs} \cite{Rasmussen_GPM_book}. 
Then, we use this surrogate model to fit the cost function and compute several scores to assess its reliability. 
For each number of samples $\Nsamples$ and for each figure of merit, we thus gather a set of scores equal to the number of training sets used; we then take the median values of those, which is a statistic that is more robust against outliers than the mean. 
In what follows, we examine in detail the results for the LiH ground state search and for the unweighted MaxCut problems.

\begin{figure*}
  \centering
  \includegraphics[width=\textwidth]{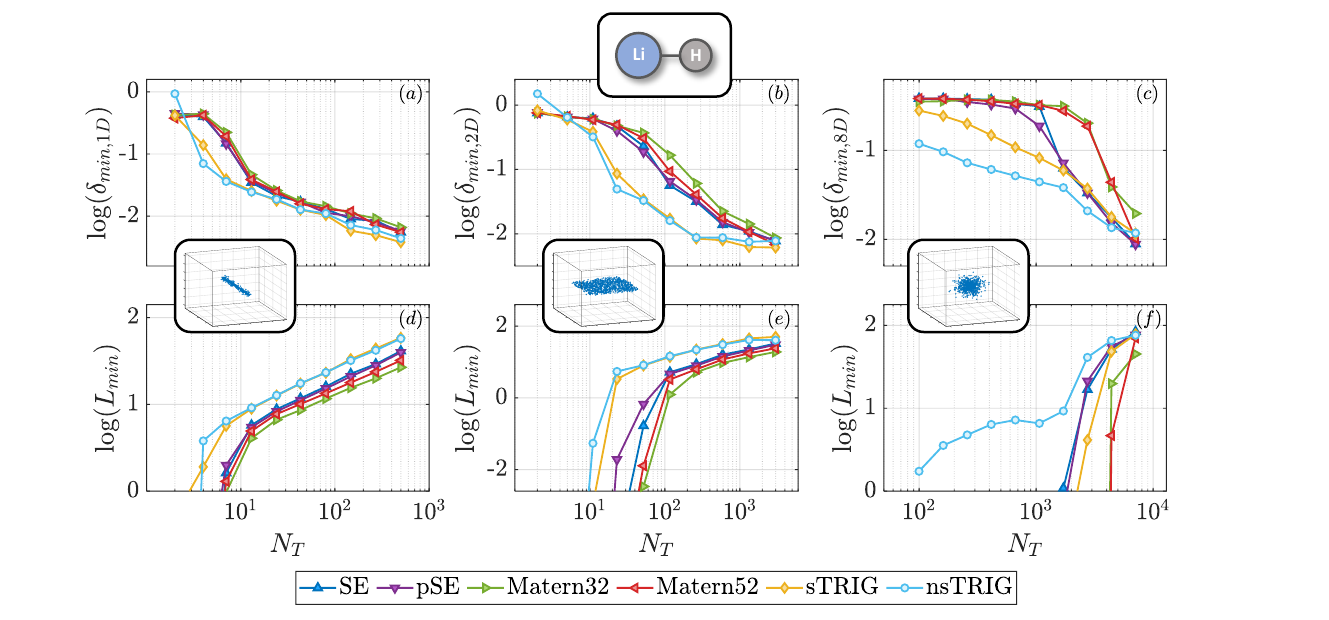}
  \caption{
  Assessing the GPM capability of predicting the correct cost function value at the exact global minimum point, using different kernels and for the LiH ground state search problem. Different columns refer to the different subspaces described in Sec.~\ref{sec:KernelComparison:HamiltoniansAndSampling}. 
  Top row: distance between GPM-predicted and true cost function values at the global minimum, $\delta_{min, \subspacelabel}$ from Eq.~\eqref{eq:KernelComparison:FiguresOfMerit:distance_predictions_minimum}, for different training set sizes $\Nsamples$. 
  Bottom row: likelihood at the minimum, $\likelihood_{min}$ from Eq.~\eqref{eq:KernelComparison:FiguresOfMerit:likelihood_minimum}, for different training set sizes $\Nsamples$. 
  }
  \label{fig:KernelComparison:LikelihoodMinimum_LiH}
\end{figure*}

\subsection{Minimum point prediction}
\label{sec:KernelComparison:MinimumPoint}
One crucial figure of merit for GPMs employed in VQAs is the ability of the GPM to locate the minimum point of the noisy cost function correctly, within a subspace (or subset of points) $\subspacelabel$ of the entire parameter space. 
We, therefore, define a first figure of merit as the \textit{distance between the location of the exact minimum and the GPM-predicted minimum location, within $\subspacelabel$}, 
\begin{equation}
    \mindist{\subspacelabel} = || \vecx_{min, \subspacelabel}^{\noiseless} - \vecx_{min, \subspacelabel}^{\GPM} ||_2 \; ,
    \label{eq:KernelComparison:FiguresOfMerit:mindist}
\end{equation}
where $\vecx_{min, \subspacelabel}^{\noiseless} = \argmin_{\vecx \in \subspacelabel} \cf(\vecx)$ is the global minimum point of the cost function within $\subspacelabel$, 
while $\vecx_{min, \subspacelabel}^{\GPM} = \argmin_{\vecx \in \subspacelabel} \ftestmean(\vecx)$ is the minimum point of the GPM, computed in the subspace $\subspacelabel$.
The minimum search over the GPM-predicted manifold is done by starting from a neighborhood of the global minimum point to avoid possible local minima far away from it.

Figure~\ref{fig:KernelComparison:MinimumPoint} compares the different kernels according to the metric $\mindist{\subspacelabel}$ for increasing training set size $\Nsamples$. 
Panels in different columns refer to training samples belonging to the different subspaces $\subspacelabel = $ 1D, 2D and 8D, introduced in Sec.~\ref{sec:KernelComparison:HamiltoniansAndSampling}, while the panel rows correspond to the LiH ground state search (top) and the unweighted MaxCut problem (bottom). 
As expected, the distance between the GPM-predicted and exact minimum decreases monotonically with more samples in all cases; however, trigonometric kernels generally require fewer data points to reach a given precision compared to other kernels. This is the consequence of putting more information about the cost function structure directly in the kernel. 
An exception to this conclusion is Fig.~\ref{fig:KernelComparison:MinimumPoint}(f), where the non-stationary trigonometric kernel is performing worse than all the other ones, although we note that in this case, all kernels give particularly bad scores, especially for a larger number of samples. 
We will see in the next section that the training set is, in this case, particularly hard to handle within the GPM framework because of the strongly non-Gaussian shot noise for the unweighted MaxCut cost function close to the global minimum.

\begin{figure*}[ht]
  \centering
  \includegraphics[width=\textwidth]{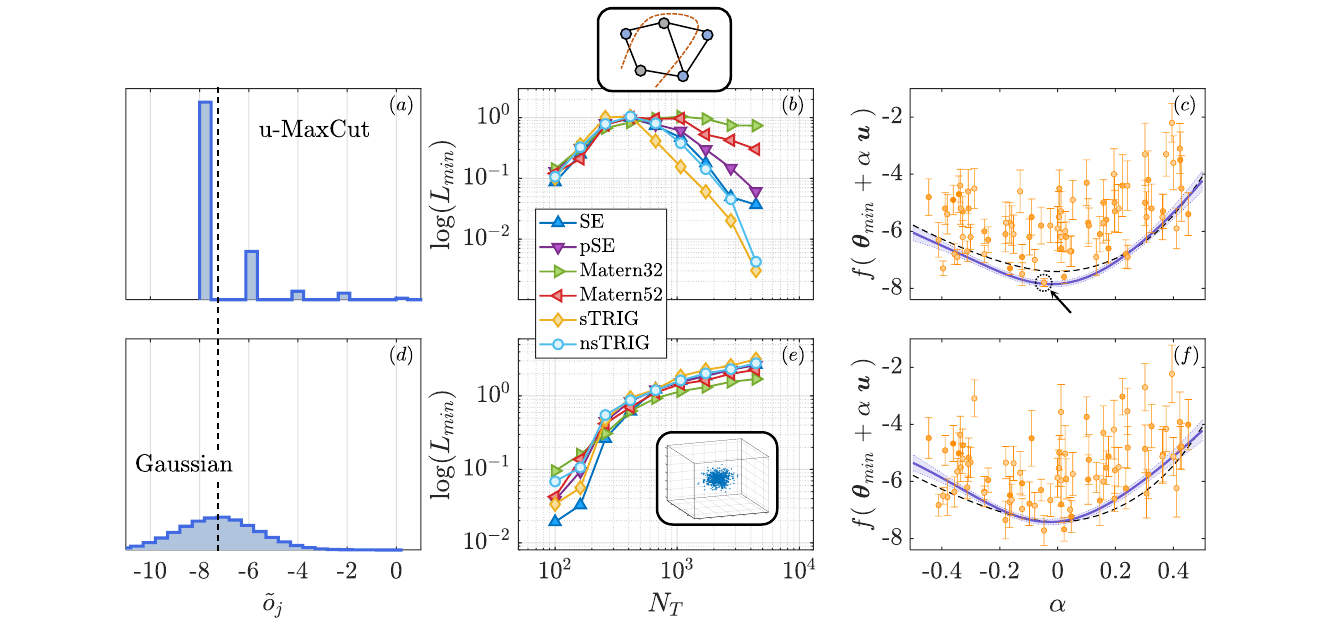}
  \caption{
  Analysis of the failure of GPM fits for the unweighted MaxCut problem for samples close to the exact global minimum point $\vecx_{min}$. 
  The first row refers to projective measurements as coming from a real device, while the second row shows what happens when considering Gaussian noise. 
  Panels (a),(d): distribution of projective measurements, $\tilde{o}_j = \sum_i o_{ij}$ (see Eq.s \eqref{eq:VQA:costfunction} and \eqref{eq:VQA:finiteSampling}), at $\vecx_{min}$ as in a real device and according to the artificially-transformed observations, respectively (mean indicated by the black dashed line). From this, the cost function is estimated by 
  $\sum_{j=1}^\s \tilde{o}_j / \s$. 
  Panels (b),(e): $\likelihood_{min}$ as a function of the training set size $\Nsamples$. 
  Right column: instance of an individual fit along a random line in parameter space centered at $\vecx_{min}$ with direction $\textbf{u}$, for realistic (c) and artificial Gaussian (f) noise; the GPM prediction (blue solid line) with associated error bar (shaded area around) is plotted against the exact cost function (black dashed line); the orange dots with error bars are cost function estimations (the 80 closest points to the line out of the 2748 overall training points), where less transparent means the point lies closer to the line. 
  The black arrow in (c) points at one problematic sample. 
  }
  \label{fig:KernelComparison:uwMaxCut_LikeMinAndProbDistributions}
\end{figure*}

\subsection{Minimum value prediction}
\label{sec:KernelComparison:Minimum}
Even if the GPM locates the global minimum point well, the GPM prediction of the actual function value at the minimum might still be incorrect or have a large uncertainty.
To quantify this, we compute the posterior distribution of cost function values $y$,
at the global minimum point $\vecx_{min}^{\noiseless}$ of the noiseless cost function from Eq.~\eqref{eq:VQA:costfunction}. That is, we compute the (normally distributed) conditional probability density $p_{f_*}(y | \Xtest, \Xtrain, \mathbf{y})$ from Eq.~\eqref{eq:GPMposterior} for the single test point $\Xtest = \vecx_{min}^{\noiseless}$, conditioned on the training set of parameters $\Xtrain$ and observations $\mathbf{y}$.
Then, the \textit{likelihood} that the GPM predicts the \textit{exact global minimum value} 
$y_{min} = \cf( \vecx_{min}^{\noiseless} )$ at $\vecx_{min}^{\noiseless}$ is 
\begin{equation}
    \likelihood_{min} = p_{f_*}(y_{min} | \Xtest = \vecx_{min}, \Xtrain, \mathbf{y}) \; .
    \label{eq:KernelComparison:FiguresOfMerit:likelihood_minimum}
\end{equation}
The concept of likelihood is illustrated graphically in Fig.~\ref{fig:visualizeGPM}(b): 
it has a high value when there is good agreement between mean prediction and true value \textit{and} a small uncertainty; if the standard deviation of the posterior distribution is instead large, the likelihood decreases. 
Similarly, a prediction close to the true value but with a very small standard deviation that does not reach the true function value yields a very small likelihood. 

To provide better distinguish between the two sources of low likelihoods, we complement the analysis by also evaluating the \textit{difference between GPM-predicted and true cost function values at the global minimum},
\begin{equation}
    \delta_{min, \subspacelabel} = 
    \frac{ |y_{min}^{\GPM} - y_{min}| }{ y_{max, \subspacelabel} - y_{min} } \; ,
    \label{eq:KernelComparison:FiguresOfMerit:distance_predictions_minimum}
\end{equation}
where $y_{min}^{\GPM}$ is the mean of the (Gaussian) posterior distribution $p_{f_*}(y | \Xtest = \vecx_{min}, \Xtrain, \mathbf{y})$ introduced above, 
while $y_{max, \subspacelabel}$ is the cost function maximum within subspace $\subspacelabel$.

Let us first discuss the results for the quantum chemistry problem. Figure~\ref{fig:KernelComparison:LikelihoodMinimum_LiH} shows the behavior of both $\delta_{min}$ (top row) and $\likelihood_{min}$ (bottom row), under the same conditions employed for the results discussed in the previous section. Each column refers to the different sampling strategies outlined in Sec.~\ref{sec:KernelComparison:HamiltoniansAndSampling}. 
In all scenarios, the trigonometric kernels provide a better estimate of the cost function value at the global minimum point, in agreement with the previous results in Fig.~\ref{fig:KernelComparison:MinimumPoint}(a)-(c).

The results for the unweighted MaxCut are instead more surprising. We find that while the $\likelihood_{min}$ initially increases with growing training set size, it starts to drop again for larger $\Nsamples$. This happens for all kernels, as can be seen in Fig.~\ref{fig:KernelComparison:uwMaxCut_LikeMinAndProbDistributions}(b) for training data lying in a hyperball centered at the global minimum point (sampling strategy 3 in Sec.~\ref{sec:KernelComparison:HamiltoniansAndSampling}). 
The results for $\delta_{min, 8D}$ exhibit similar issues, as curves stay constant or increase for large $\Nsamples$ (not shown).

This behavior is counter-intuitive because the surrogate model quality is expected to monotonically increase with the number of samples. It can be explained by looking explicitly at individual fits. 
Fig.~\ref{fig:KernelComparison:uwMaxCut_LikeMinAndProbDistributions}(c) displays the surrogate model fit and the exact cost function along a random line passing through the global minimum point in the same sampling conditions of panel (b). The orange dots are the closest 80 samples to this line out of a total of 2748 points. 
We immediately see that the fit significantly underestimates the true cost function values, especially at the global minimum ($\alpha = 0$).

\begin{figure*}
  \centering
  \includegraphics[width=\textwidth]{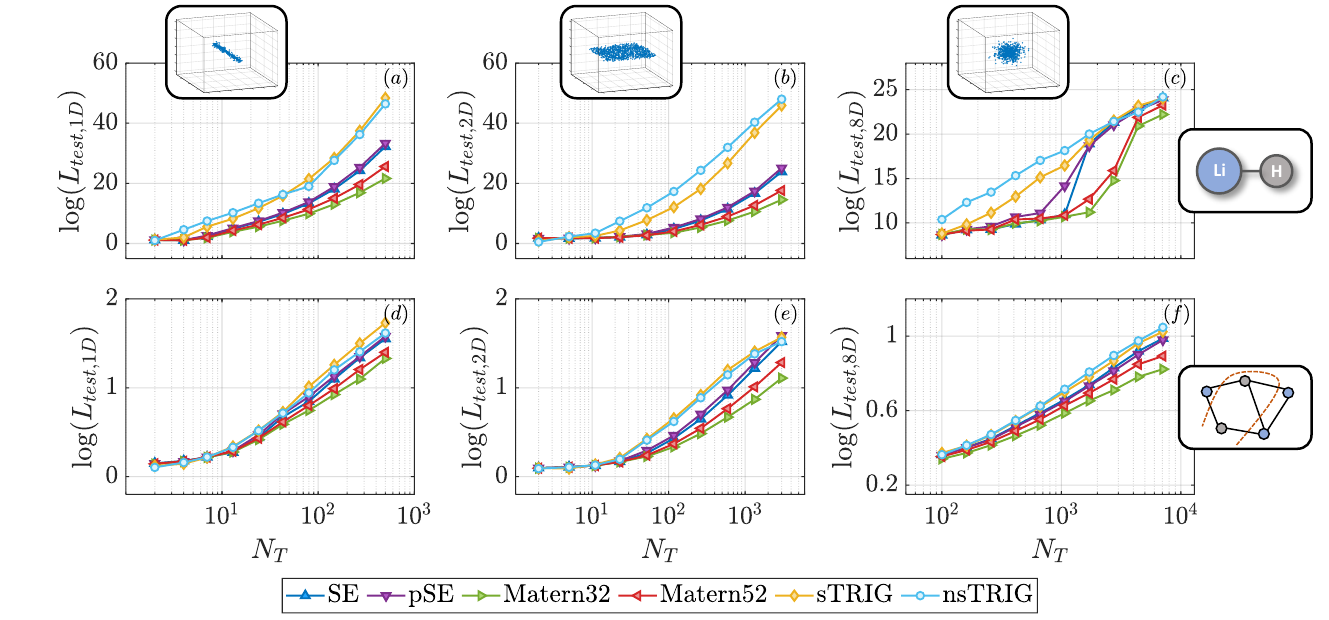}
  \caption{
  GPM fitting quality as measured by mean likelihood of all samples in the test set $\subspacelabel$,
  using different kernels and for the LiH ground state search (top) and the unweighted MaxCut (bottom) problems. 
  One point for a given $\Nsamples$ is the median of the average test likelihoods $\likelihood_{test, \subspacelabel}$, computed using 100 different training sets of size $\Nsamples$. 
  Each column refers to training data taken from the different subspaces described in Sec.~\ref{sec:KernelComparison:HamiltoniansAndSampling}. 
  }
  \label{fig:KernelComparison:LikelihoodTest}
\end{figure*}

This behavior appears to be caused by data points occurring close to the minimum, which have very low values and very small error bars that do not include the exact cost function value. Such anomalous points result from a breakdown of the Gaussianity assumption on the data. That is, close to the minimum, the discretized nature of the shot noise becomes more apparent, as illustrated in Fig.~\ref{fig:KernelComparison:uwMaxCut_LikeMinAndProbDistributions}(a), where the projective measurement distribution at the global minimum is shown. 
It can happen that, taking $s=20$ shots, projective measurements return almost always the lowest value in the distribution, leading to a mean lower than the true one and, crucially, to a very small variance. 
The GPM then forces the fit to pass through or very near this point, leading to the observed underestimation of the true function value. 
This problem worsens with increasing training set size (but keeping $s=20$ shots) since the probability of getting at least one of these problematic samples increases. For points further away from the minimum, fluctuations around the true mean function value are symmetric, leading to no bias, and the distribution of measurement outcomes is not imbalanced towards a single value.

To verify that the non-gaussianity and discreteness of measurement outcomes are indeed the reason for the GPM failure, we repeated the same inference in a setting where we made projective measurement noise to be Gaussian by hand, using the following procedure: 
Given a point in parameter space, we compute the exact mean and standard deviation of the projective measurement distribution; then, we sample from a Gaussian distribution with the same mean and standard deviation (see Fig.~\ref{fig:KernelComparison:uwMaxCut_LikeMinAndProbDistributions}(d)), using the same number of shots $\s$, to obtain new mean and standard deviation estimates. 

With these new 'artificial' cost function estimations, we repeated the inference on the very same data as before, obtaining that the likelihood of the minimum increases monotonically with increasing number of samples in the training set, as shown in Fig.~\ref{fig:KernelComparison:uwMaxCut_LikeMinAndProbDistributions}(e). 
Fig.~\ref{fig:KernelComparison:uwMaxCut_LikeMinAndProbDistributions}(f) displays the individual fit within the same conditions of Fig.~\ref{fig:KernelComparison:uwMaxCut_LikeMinAndProbDistributions}(c), but with cost function estimations from artificial Gaussian projective measurement distributions. As expected, there are no more isolated points with tiny error bars, and the fit is much more reliable. 

We remark that the previous issue specifically concerns cost functions made of single Pauli strings: when the cost function contains multiple non-commuting observables, like in quantum chemistry and quantum many-body ground state search, their sampled estimates are subject to quantum fluctuations; as a consequence, the projective measurement distribution for the cost function is much more similar to that of Gaussian and much less skewed.

\subsection{Overall fit quality}
\label{sec:KernelComparison:TestSet}
In this section we assess the goodness of a GPM fit over the whole space covered by the training data instead of merely evaluating quality near the optimum value as in the previous sections. 
To this end, we randomly select a test set of points from the same subspace $\subspacelabel$ that the training samples belong to and define the \textit{test likelihood} $\likelihood_{test, \subspacelabel}$ as the mean likelihood of each individual test point. 
Fig.~\ref{fig:KernelComparison:LikelihoodTest} shows how the test likelihood monotonically increases for increasing training set size $\Nsamples$. 
In both problems, the trigonometric kernels provide a better overall cost function fit than the other kernels. This is especially true for the LiH ground state search (top row), where we observe much larger test likelihoods than for the unweighted MaxCut problem (bottom row).

\section{RotoGP: a GPM-enhanced sequential line search optimizer}
\label{sec:ROTOGP}
\begin{figure*}
  \centering
  \includegraphics[width=\textwidth]{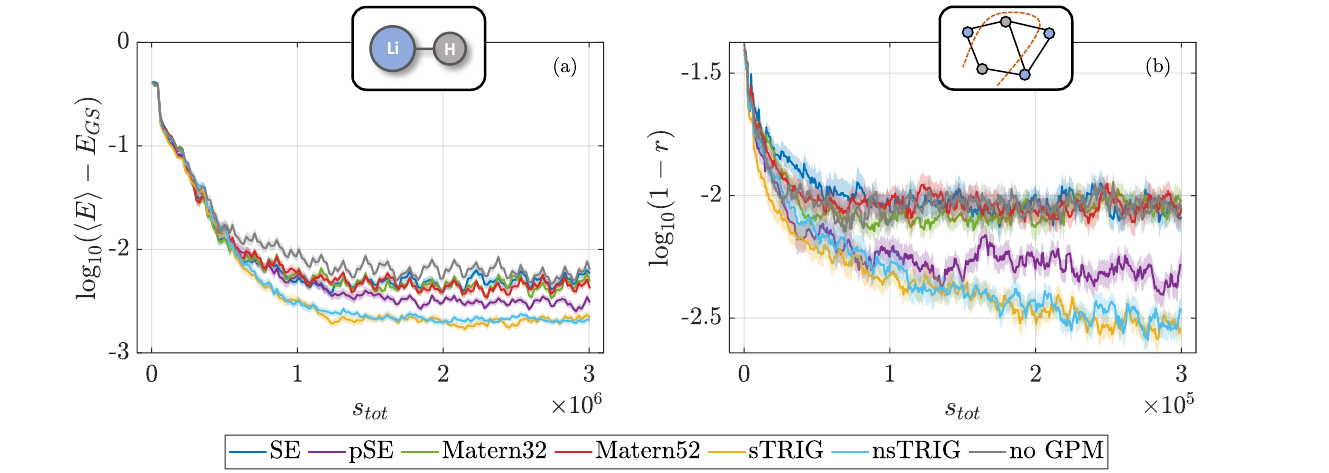}
  \caption{Average optimization trajectories with RotoGP when minimizing (a) the ground state energy of the LiH molecule and (b) the cost function of the unweighted MaxCut problem, using GPM equipped with different kernels and without GPM. 
  Each line is the average over 100 runs from different initial points and, for the combinatorial problem, different graphs. 
  The x-axes show the total number of shots used during the optimization, $\s_{tot}$. 
  The y-axis shows (a) the distance from the ground state energy and (b) the deviation of the approximation ratio from one (corresponding to the exact solution). 
  }
  \label{fig:ROTO_optimization}
\end{figure*}
In this section, we investigate how a GPM can improve convergence when employed in an actual variational quantum optimization. To this end, we equip a standard sequential line search optimizer~\cite{NFT_PRR20, SLS_Parrish_arXiv19, Rotosolve_Quantum21} with a GPM to improve cost function estimates and benchmark the optimization performance using different kernels. 
In particular, we combine the RotoSolve optimizer introduced in Ref.~\cite{Rotosolve_Quantum21} with a GPM and call this new optimizer RotoGP. 
Its strategy is to update one parameter per iteration while keeping all the others fixed. 
To do so, one samples along a given coordinate line at specific points and then exploits an analytical formula to compute the global minimum along that line. 
The original RotoSolve was designed for cost functions having a single frequency along each coordinate line~\cite{Rotosolve_Quantum21}; however, it can be generalized to cases with multiple frequencies $\nfROTO$, at the expense of increasing the number of samples to take along the line to $2\nfROTO + 1$~\cite{GeneralizedParShiftRule_Quantum22}. 
A good surrogate model can now help mitigate the effect of noise, providing more accurate estimates (in principle using information from all samples in the training set, also samples that are \textit{not} locatd on the current coordinate line) and improving optimization convergence. 

We employ RotoGP for both problems considered before: ground state search of the LiH molecule 
and solving the unweighted MaxCut problem with QAOA. We then compare its performance using different GPM kernels, as well as the case without GPM (the bare RotoSolve). 
Fig.~\ref{fig:ROTO_optimization} shows the optimization progress as a function of the shots used. 
For the quantum chemistry problem, we plot the distance between the expectation value of the energy and the true ground state energy $E_{GS}$, while for the unweighted MaxCut, we observe how much the approximation ratio, defined in Eq.~\eqref{eq:approxratio} of Appendix~\ref{app:uwMaxCut_intro}, deviates from one, corresponding to the exact solution. 
In both cases, the trigonometric kernels improve upon the performance of the bare RotoSolve without GPM: we find both higher accuracy and faster convergence. 
This is also true for the pSE kernel, but the improvement is smaller than the one corresponding to the trigonometric kernels. 
On the other hand, all the other kernels provide similar results to the case without GPM, proving that a good choice of the kernel is crucial. 

Finally, we provide some details on how the optimization runs are carried out.
Following the results discussed in Appendix~\ref{app:fewfreqs}, we use $\nfROTO = 2$ frequencies on all coordinate lines for the chemistry problem and $\nfROTO = 4$ and $\nfROTO = 1$ for the odd and even coordinate lines of the combinatorial problem, respectively. 
To construct a trigonometric line kernel, we therefore set the frequency cutoff $n_f$ in Eqs.~\eqref{eq:kernel:TRIG} and~\eqref{eq:kernel:TRIGstat} to be equal to $\nfROTO$ along that coordinate line, as already discussed in Sec.~\ref{sec:KernelComparison:HamiltoniansAndSampling}. 
The kernel hyperparameters are optimized at each iteration step, that is, after sampling along a coordinate line and before estimating the new minimum point candidate, by maximizing the logarithm of the likelihood defined in Eq.~\eqref{eq:GPMlikelihoodHPs}. 
To mitigate the computational time that would increase with the training set size, we only consider a subset of at most 300 samples for the hyperparameter optimization, picking at each iteration only the closest to the current optimization line.
We fix the number of shots to measure the expectation value of a Pauli string operator to $\s = 100$. 
Finally, for the quantum chemistry problem, we start the optimization from a random point in parameter space rather than initializing all parameters to zero, as is commonly done for the UCCSD Ansatz. This is to make the optimization problem more complex and better see the difference in performance using the different kernels. 

\section{Conclusions}
In this work, we have explored the role of kernel choice in Gaussian Process Models (GPMs) for fitting cost functions in Variational Quantum Algorithms (VQA). We have introduced trigonometric kernels, a novel class of kernels tailored for periodic VQA cost functions. Such periodic cost functions appear naturally in VQAs and, in many cases, contain only surprisingly few frequencies. Trigonometric kernels demonstrate superior GPM fitting performance on static benchmarks such as the distance-to-minimum, showcasing their capacity to enhance VQA cost function modeling. 

The advantages of trigonometric kernels are most apparent when only very few and noisy samples are available. This makes them particularly suitable for quantum hardware where measurements are costly, such as Rydberg atom and ion-trap platforms, as opposed to faster super-conducting hardware. However, in such few-shots limits, care must be taken regarding the distribution of the shot noise, which might not be Gaussian but rather follow a generalized Bernoulli distribution~\cite{McClean_TheoryVQE_NJP16}, as we have encountered for the case of the MaxCut QAOA problem. Adapting the GPM framework to correctly incorporate such distributions could be an interesting future research direction.

Trigonometric kernels are able to utilize data from the entire parameter space rather than relying on nearby points, which could explain their strong performance. This property suggests they may especially excel in global optimization algorithms. Additionally, the relative gains of trigonometric kernels appear problem-dependent. Overall, this invites future research to identify broader classes of optimization algorithms and VQA problems suited to particular kernels.

We have also evaluated kernel performance in an actual (simulated) optimization setting, using the newly developed RotoGP, a new optimization algorithm based on RotoSolve~\cite{Rotosolve_Quantum21}, that incorporates a GPM to mitigate noise during optimization. Trigonometric kernels proved particularly effective in this setting, whereas other standard kernels performed significantly worse or, in some cases, even failed to improve over the base RotoSolve algorithm.

Overall, our findings emphasize the importance of careful kernel selection in quantum optimization algorithm design. In particular, trigonometric kernels provide a promising avenue for enhancing performance in challenging optimization tasks on present-day quantum hardware.

\section{Acknowledgements}
This work was funded through the Austrian Research Promotion Agency (FFG) contract 884471 (ELQO), and the European High-Performance Computing Joint Undertaking (JU) under grant agreement No 101018180 HPCQS.
The computational results presented here have been achieved (in part) using the LEO HPC infrastructure of the University of Innsbruck.

\bibliographystyle{unsrtnat}
\bibliography{biblioGPM}

\clearpage

\onecolumn
\appendix

\section{Derivation of trigonometric kernels assuming exponentially-decaying weights}
\label{app:trigDetails}
We provide here the derivation of the trigonometric kernels, assuming that the weights $\w$ decay exponentially with the frequency number after a given threshold $n_f$. This provides even more flexible kernels than the ones derived in the main text. 
Let us start with a generic trigonometric kernel with all free hyperparameters for any frequency, 
\begin{equation}
    k(\x, \x') = \gamma_0 + \sum_{n=1}^\infty \Big[ \gamma_n \cos (n\x) \cos (n\x') 
    + \delta_n \sin (n\x) \sin (n\x') \Big] \; .
\end{equation}
We then set the prior covariance matrix diagonal as $[\Sigma_p]_{ii} \simeq \exp(-\beta n_i)$, with $n_i$ the frequency of the $i^{th}$ basis function of Eq.~\eqref{eq:basisfunctionsvector}, and doing this only for those components $i$ for which $n_i > n_f$, with $n_f$ the threshold frequency. The newly introduced hyperparameter $\beta$ controls the decay rate. Then, we can evaluate 
\begin{align}
\begin{split}
    \sum_{n=n_f + 1}^\infty e^{-\beta n} \cos \left( n (\x-\x') \right) &= 
    \sum_{n=1}^\infty e^{-\beta n} \cos \left( n (\x-\x') \right) 
    - \sum_{n= 1}^{n_f} e^{-\beta n} \cos \left( n (\x-\x') \right) = \\
    &= \textrm{Re} \left\{ \frac{1}{e^{\beta + i (\x-\x')} - 1} \right\} 
    - \sum_{n= 1}^{n_f} e^{-\beta n} \cos \left( n (\x-\x') \right) \; .
\end{split}
\end{align}
The second factor can be neglected because each $e^{-\beta n}$, $n \leq n_f$, can be absorbed by its corresponding $\gamma_n$ and $\delta_n$ hyperparameters. We thus arrive at an extended version of the non-stationary trigonometric kernel, 
\begin{align}
\begin{split}
    k_T^{(ns, ext)}(\x, \x') = \Bigg\{\sum_{n=1}^{n_f} \Big[ \gamma_n \cos (n\x) \cos (n\x')                            + \delta_n \sin (n\x) \sin (n\x') \Big] 
                + \gamma_0 + \alpha \real \left[ \frac{e^\beta - 1}{e^{\beta + i (\x-\x')} - 1} \right] \Bigg\} / \Gamma,
\end{split}
\end{align}
where $\alpha$ is a new hyperparameter governing the amplitude of the exponentially decaying term. 
The normalization is here $\Gamma = \gamma_0 + \frac{1}{2} \sum_{n=1}^{n_f} \left( \gamma_n + \delta_n \right) + \alpha $. 
The corresponding stationary version of this kernel takes the form 
\begin{equation}
    k_{T}^{(s, ext)}(\distx) = \Bigg\{ \gamma_0 + \sum_{n=1}^{n_f} \gamma_n \cos \left( n \distx \right) + + \alpha \real \left[ \frac{e^\beta - 1}{e^{\beta + i \distx} - 1} \right] \Bigg\} / \Gamma.
\end{equation}

\section{LiH ground state with UCCSD Ansatz}
\label{app:LiH_intro}
The first VQA problem we focus on is the ground state calculation of the LiH molecule at an interatomic distance $d = 1.3\AA$. 
This problem has been widely studied in the VQA-related literature for different $d$~\cite{Kandala_Nature17, Roos_Chemistry_PRX18, NFT_PRR20}. 
Under the Born-Oppenheimer approximation, we first compute a molecular orbital basis using a linear combination of Slater-Type (atomic) Orbitals, each represented as a linear combination of six Gaussian functions (STO-6G). 
In practice, this is done using the PySCF package~\cite{PySCF}. 
This leads to the Hamiltonian in the second quantization, 
\begin{equation}
    H = \sum_{pq} h_{pq} c_p^\dagger c_q + \sum_{pqrs} h_{pqrs} c_p^\dagger c_q^\dagger c_r c_s \; ,
    \label{app:LiH_intro:Hfermions}
\end{equation}
where the one-body and two-body coefficients $h_{ab}, h_{abcd}$ are computed from the molecular orbitals found in the previous step. 

The LiH molecule has 4 electrons distributed among 6 molecular orbitals. 
However, to represent the ground state, one can assume that two electrons are frozen in the lowest energetic molecular orbital while the other two do not reach the two highest energetic orbitals. 
Therefore, the problem can be simplified to two electrons placed in an active space of 3 molecular orbitals~\cite{Roos_Chemistry_PRX18}. 

Since we cannot enforce fermionic statistics on the quantum hardware, we proceed by applying the Bravyi-Kitaev mapping to cast Eq.~\eqref{app:LiH_intro:Hfermions} to a spin Hamiltonian, 
\begin{equation}
    H_{spin} = \sum_\alpha g_\alpha P_\alpha \; ,
    \label{app:LiH_intro:Hspins}
\end{equation}
where $P_\alpha$ is a tensor product of Pauli operators, and $g_\alpha$ is a coefficient. 

As a quantum circuit, we use one layer of Unitary Coupled-Cluster Ansatz truncated to Single and Double excitations (UCCSD), 
\begin{equation}
    U(\vecx) = 
    \prod_{p>r} e^{\x_{pr}( c_p^\dagger c_r - H.c. )}
    \prod_{p>q>r>s} e^{\x_{pqrs}( c_p^\dagger c_q^\dagger c_r c_s - H.c. )} \; ,
    \label{app:LiH_intro:UCCSDfermions}
\end{equation}
where $\vecx$ is the parameters vector and $r,s$ and $p,q$ run over the occupied and unoccupied molecular orbitals, respectively. 
The idea behind this Ansatz is to dress the Hartree-Fock ground state variationally with one-body and two-body excitations. Therefore, the system is always initialized to the Hartree-Fock ground state before executing the quantum circuit. 
The Ansatz in Eq.~\eqref{app:LiH_intro:UCCSDfermions} is then also expressed in the spin picture by means of the Bravyi-Kitaev mapping, leading to an evolution operator with several Pauli strings in the exponent, each multiplied by a parameter. 
We group the Pauli strings that commute with each other and assign them the same parameter. 
The circuit and Hamiltonian definition in spin space are implemented in practice using Qiskit~\cite{qiskit} and Qiskit-Nature~\cite{qiskit_nature}. 
Eventually, the circuit involves 6 qubits and 8 parameters to optimize. 
The number of pair-wise non-commuting Pauli strings in Eq.~\eqref{app:LiH_intro:Hspins} is 38. Thus, a round of measurements requires $s = 38 s_0$ shots, where $s_0$ is the number of shots used to estimate the expectation value of a single Pauli string.

\section{MaxCut problem with QAOA}
\label{app:uwMaxCut_intro}
Given an undirected graph with $N$ vertices and random (possibly weighted) connectivity among them, the MaxCut problem consists of finding a bipartition of these vertices such that the (weighted) sum of the edges broken by the bipartition is maximized. 
The cost function for this problem can be written as 
$\cf(\boldsymbol{z}) = \sum_{ij} w_{ij} (1 - z_i z_j) / 2$, 
where $z_i = \pm 1$ identifies the bipartition of the $i^{th}$ vertex and $w_{ij}$ is the weight associated to the edge connecting vertices $i$ and $j$. If $w_{ij} = 1$ for all $i, j$, one speaks about \textit{unweighted} MaxCut. 

The MaxCut problem can be tackled with VQAs by lifting each $z_i$ classical variable to a corresponding $\sz_i$ Pauli operator of quantum mechanics. The problem can thus be reformulated as finding the ground state of the following \textit{problem} Hamiltonian 
\begin{equation}
    \widehat{H}_C = \sum_{ij} w_{ij} \sz_i \sz_j \; ,
    \label{eq:HMaxCut}
\end{equation}
where anti-aligned spins give a negative contribution weighted by their corresponding edge weight. 
To solve the problem with QAOA, one introduces the so-called \textit{mixing} Hamiltonian, 
$\widehat{H}_B = \sum_{i=1}^N \sx_i$, 
which does not commute with the problem Hamiltonian, in the same spirit of quantum annealing. 
The idea is then to choose the initial state as the ground state of $\widehat{H}_B$, 
$\ket{\psi_0} = \spindown_x^{\otimes N}$, 
that is in a superposition of all the possible solutions of the problem, and then apply the problem and mixing Hamiltonian alternately as 
\begin{equation}
    \UU(\vbeta, \vgamma) = \prod_{j=1}^p e^{-i\beta_j \widehat{H}_B} e^{-i\gamma_j \widehat{H}_C} \; ,
\end{equation}
where $\vecx = (\vbeta, \vgamma)$ is the set of circuit parameters, $p$ is the number of layers, and the product is ordered so that the $p^{th}$ term is applied last. 
The cost function to minimize is thus 
\begin{equation}
    \cf(\vecx) = \mean{ \psi_0 | \UU^\dagger(\vecx) \widehat{H}_C \UU(\vecx) | \psi_0 } \; ,
\end{equation}
that is the energy of the variational state on the problem Hamiltonian. 
The circuit, once optimized, should shape the initial wavefunction so to have large amplitudes for the z-basis states corresponding to the solutions of the problem. Then, it suffices to sample one of these states at least once to accomplish the task. 
To evaluate the goodness of a variational solution, let us define the expectation value of the number of edges cut as 
$F(\vecx) = \sum_{ij} w_{ij} \mean{ \psi_f(\vecx) | (\id - \sz_i \sz_j) | \psi_f(\vecx) } / 2$, where 
$\ket{\psi_f(\vecx)} = \UU(\vecx) \ket{\psi_0}$. 
Then, a typical figure of merit is the so-called approximation ratio, \ie the ratio between $F(\vecx)$ and $F_{max}$, where the latter is the number of edges cut for the bipartition that solves the problem exactly. 
For the sake of comparing different optimization runs for a given circuit, as we do in Sec.~\ref{sec:ROTOGP}, it is convenient to define a \textit{modified approximation ratio}, where one compares the variational solution to the best achievable one, using the given quantum circuit, 
\begin{equation}
    r = \frac{F(\vecx)}{F(\vecx_{best})} \; ,
    \label{eq:approxratio}
\end{equation}
where $\vecx_{best}$ is the circuit minimum point and $r \leq 1$.

\section{Frequency analysis of cost functions}
\label{app:fewfreqs}
\begin{figure}
  \hspace{1.5cm}
  \includegraphics[width=0.7\textwidth]{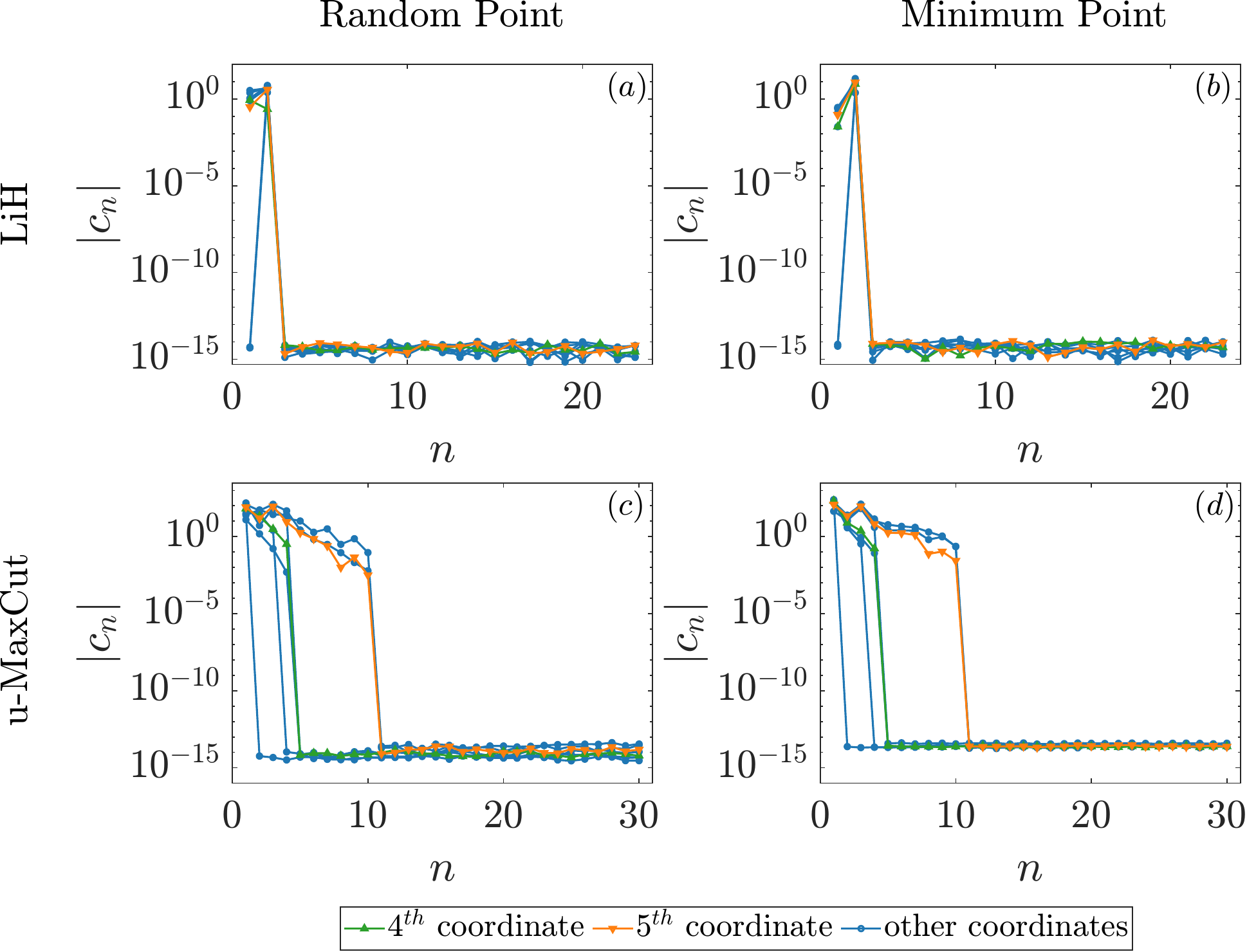}
  \caption{Absolute value of the cost function Fourier coefficients when scanning along single coordinate lines, all passing through the same point. The latter point is taken to be random (left column) and the exact minimum point (right column). 
  The top row refers to the LiH problem with UCCSD Ansatz; the bottom row corresponds instead to the unweighted MaxCut problem with QAOA (average over 50 random instances). 
  }
  \label{fig:fewfreqs}
\end{figure}
We give here empirical evidence on the fact that the cost functions under study in this work are characterized by much fewer frequencies than a priori expected, in line with previous findings on MaxCut problems with QAOA~\cite{FastGradient_Bittel_arXiv22}. 
In our analysis, we consider the two problems under study in this work, and look at the Fourier spectrum of their cost functions along coordinate lines, centered both at a random point in parameter space and at the global minimum point. 

Fig.~\ref{fig:fewfreqs} shows the absolute value of the Fourier coefficients in the different settings. 
The LiH cost function (panels (a)-(b)) is exactly described by the lowest two frequencies. 
They can be equally important, in general. For example, from panel (a) we see that the first frequency is more relevant for the $4^{th}$ coordinate, while the second frequency is more important for the $5^{th}$ coordinate. 
However, when focusing on coordinate lines passing through the global minimum point (panel (b)), the second frequency becomes almost the only one that matters. 

The unweighted MaxCut cost function is also characterized by its lowest frequencies only, as evident from panels (c)-(d).
For the coordinate lines corresponding to the mixing Hamiltonian (even coordinates), only the first frequency really matters, while the others decay exponentially fast. 
For parameters corresponding to the problem Hamiltonian, we observe instead up to four frequencies relevant to characterize the cost function along the coordinate line, with the subsequent ones again decaying exponentially with the frequency number. 
Notice that, given the stochastic nature of the MaxCut problems, we have averaged the results over 50 different graph instances. 

\begin{figure}
  \centering
  \includegraphics[width=0.45\textwidth]{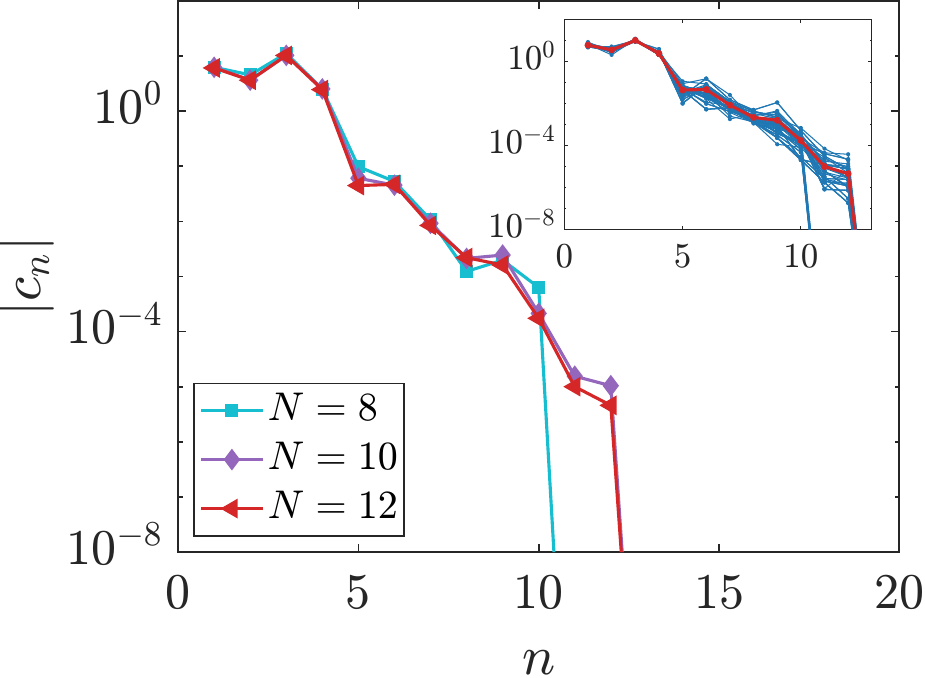}
  \caption{Same analysis as in Fig.~\ref{fig:fewfreqs}, but for unweighted MaxCut problems of different sizes, \ie containing $N = 8, 10, 12$ vertices. 
  The plot shows the average Fourier spectrum for the $5^{th}$ coordinate lines only, passing through random points (average over 50 instances). 
  We note that increasing $N$ does not change the exponential suppression of the frequencies. 
  Inset: all the 50 different sets of absolute value Fourier coefficients (blue) and their average (red) for $N = 12$ vertices.}
  \label{fig:fewfreqs_scaling}
\end{figure}
To verify that this is not just related to the small size of the problem, we also checked the MaxCut case at different number of vertices. 
The results in Fig.~\ref{fig:fewfreqs_scaling} show evidence that the exponential suppression of the relevant frequencies persists with increasing number of vertices $N$ in the graphs.
The inset gives an idea of the fluctuations of the magnitude of these coefficients, for the case with 12 vertices.

\end{document}